\documentclass[aps,preprint,groupedaddress]{revtex4}
\usepackage{graphicx}

\begin{document}

\title{Orbital and Spin contributions to the $g$-tensors 
in metal nanoparticles}


\author{A.~Cehovin}
\affiliation{Division of Solid State Theory, Department of Physics,
Lund University, SE-223 62 Lund, Sweden}
\author{C.M.~Canali}
\affiliation{Department of Technology, Kalmar University, 391 82 Kalmar,
Sweden,
and Division of Solid State Theory, Department of Physics,
Lund University, SE-223 62 Lund, Sweden}
\author{A.H.~MacDonald}
\affiliation{Department of Physics, University of Texas at Austin,
Austin TX 78712}


\date{\today}

\begin{abstract}

We present a theoretical study of the mesoscopic fluctuations of $g$-tensors
in a metal nanoparticle. The calculations were performed using a 
semi-realistic tight-binding model, which contains both spin and orbital
contributions to the $g$-tensors. The results depend on the product
of the spin-orbit scattering time $\tau_{\textrm{\small so}}$ and the
mean-level spacing $\delta$, but are otherwise weakly affected
by the specific shape of a {\it generic} nanoparticle.
We find that the spin contribution to the $g$-tensors
agrees with Random Matrix Theory (RMT)
predictions. On the other hand,
in the strong
spin-orbit coupling limit $\delta \tau_{\textrm{\small so}}/\hbar \to 0$,
the orbital contribution depends
crucially on the space character of the quasi-particle wavefunctions:
it levels off at a small value for states of $d$ character but is strongly
enhanced for states of $sp$ character.
Our numerical results demonstrate that when orbital coupling to
the field is included, RMT predictions overestimate the typical   
g-factor of orbitals that have dominant d-character.  This finding 
points to a possible source of the puzzling discrepancy between theory and 
experiment.
\end{abstract}

\maketitle


\section{Introduction}

The difference in energy between the eigenvalues for the two spin states
of an isolated electron in an external magnetic field 
$B$ is $g_0 \mu_{\rm B}B$, where $\mu_{\rm B}$ is the Bohr magneton and
$g_0$ is the so-called $g$-factor for an isolated electron. If we ignore
quantum electrodynamical corrections, it follows from the one-particle
Dirac equation that $g_0=2$. 
In an infinite crystal in the absence of external magnetic fields,
time-reversal symmetry
dictates that each electron Bloch state, labeled by $\bf k$, 
is doubly degenerate (Kramer's degeneracy),
provided that the crystal has inversion symmetry\cite{kittelQTS}. 
This degeneracy is lifted by a magnetic field $B \hat B$,
and the energy difference between the two states can be expressed by means of a 
symmetric, $3\times 3\ $, ``$g$-tensor'', ${\cal G}_{\bf k}$\cite{slichter_PoMR}
\begin{equation}
\Delta \epsilon_{\bf k} = 
\mu_{\rm B}B\big(\hat B^T {\cal G}_{\bf k} \hat B \big )^{1/2}\; .
\label{eshift}
\end{equation}
The tensor ${\cal G}_{\bf k}$ is related to the total magnetic moment of the
state $\bf k$ (see below)
\begin{equation}
\mu_{\rm B} \langle \vec\mu\ \rangle_{\bf k} = 
\mu_{\rm B}  \langle \vec L\rangle_{\bf k} + 
\mu_{\rm B}  \langle  g_0 \vec S\rangle_{\bf k}\;.
\label{mmoment}
\end{equation}
Here $\mu_{\rm B} \vec L$ and $\mu_{\rm B} g_0 \vec S$ are the orbital 
and spin magnetic moments. The brackets $\langle \dots\rangle_{\bf k}$ 
denote the expectation value in the $\bf k$-state.  In the absence of
spin-orbit interactions,
the orbital angular momentum is quenched, and
only the spin magnetic moment contributes to the
level splitting. In this case the Kramer doublet is 
composed of two (opposite) pure spin
states and the tensor ${\cal G}_{\bf k}$ is isotropic,
$({\cal G}_{\bf k})_{ij} = 4\delta_{ij}$. It follows that the $g$-factor for
a magnetic field in the $z$ direction, defined as the square root of the
tensor element $({\cal G}_{\bf k})_{zz}$, is equal to the free electron $g_0$.
The effect of spin-orbit coupling  is fourfold: 
(i) because the states forming a Kramer doublet 
are no longer pure spin states, their average spin is less than $1/2$,
which tends to decrease the typical $g$-factor; (ii) the orbital angular
momentum is no longer quenched and the corresponding magnetic moment
can contribute to the level splitting, either by decreasing or increasing $g$;
(iii) the tensor structure of ${\cal G}$ is non-trivial (i.e. the response
to a magnetic field is anisotropic and does not in general lead a 
moment aligned with the field);
(iv) ${\cal G}_{\bf k}$ can vary 
strongly with ${\bf k}$. In bulk metals direct measurement of 
electron $g$-factors can be 
obtained only via conduction electron spin resonance (CESR) experiments.
CESR involves transitions between states of the electron continuum.
The two quantities measured in these experiments are the position 
of a resonance line, that is the average $g$-factor of the conduction electrons,
over the Fermi surface, $g_{\rm av}$,
and the linewidth, corresponding to the spin relaxation time 
$\tau_{\textrm{\small so}}$. 
In case of weak spin-orbit interaction the $g$-factor shift 
$\Delta g \equiv g_{\rm av} - g_0$ can be evaluated\cite{elliott1954b}
perturbatively in the spin-orbit
coupling constant $\xi$, $\Delta g \sim \xi/W$, where $W$ is the band width.
Since $W >>\xi$,  the effect of spin-orbit interaction
on bulk $g$-factors is expected to be small, even for heavy elements like gold. 
Indeed experimentally $g_{\rm av}$ in Al is essentially equal to $g_0$, while
for Au $g_{\rm av} =2.1$. When evaluated perturbatively,
the spin relaxation time $\tau_{\textrm{\small so}}$ 
can be related to $\Delta g$ and to
the ``resistivity"
relaxation time $\tau_r$ through the Elliott relation\cite{elliott1954b},
${1\over \tau_{\textrm{\small so}}}= {\Delta g^2 \over \tau_r}$.
This relation shows that scattering off impurities, surfaces, and phonons,
which determine $\tau_r$, affects indirectly
$\tau_{\textrm{\small so}}$, although the mechanism responsible for the
spin relaxation is the
spin-orbit interaction, which is essentially atomic in character.

In this paper we focus on the
$g$-factors for electrons confined inside a metal
grain of nanometer size, where the quantum energy spectrum is discrete 
and its individual quasiparticle energy levels $\epsilon_n$ can
be directly observed at low temperatures. By replacing the Bloch index $\bf k$
with the discrete quasiparticle orbital index $n$,  
equations (\ref{eshift}) and  (\ref{mmoment})
describe how a Kramer doublet splits in a magnetic field.
In contrast to the bulk case, the effects on the spin-orbit interaction  
on $g$ -- summarized in (i)-(iv) above -- 
is expected to be enhanced, since the relevant energy scale with which the
spin-orbit coupling strength $\xi$ should be compared
is not $W$ but the much smaller single-particle mean-level spacing $\delta$.
In fact quantum finite-size effects on $g$-factors in ensembles of metal 
nanoparticles  have
been investigated in the past experimentally via
CESR. (For a review see Ref.~\onlinecite{halperin1986} and references therein.) 
To the best of our knowledge though, these experiments are   
riddled with many puzzling features, which make their comparison
with proposed theories\cite{kawabata1970,buttetPRM82}
very difficult. Furthermore in none of these experiments
is there sufficient detail to extract information concerning
statistical distribution of $g$-factors (see point (iii) above ). 
On the other
hand, 
two groups\cite{salinas99,davidovicPRB00,petta2001,petta2002}, 
using tunneling spectroscopy in 
single-electron transistors, have recently succeeded in 
measuring $g$-factors of
individual quasiparticle levels of a single metal nanoparticle.
Measured $g$-factors ranged from 0.1 to 2, depending on material,
grain size and doping; $g$-factors displayed large level-to-level
fluctuations and strong dependence on the orientation of the applied
magnetic field. Clearly $g$-factors in small metal grains measured by
single-electron spectroscopy have little in common with bulk $g$-factors 
measured by CESR. 

The statistical properties and mesoscopic fluctuations
of $g$-factors
measured experimentally 
in Refs.~\onlinecite{salinas99,davidovicPRB00,petta2001,petta2002}
are well described by theoretical distributions
based on Random Matrix Theory (RMT), 
derived by two independent 
groups\cite{brouwerPRL2000, matveev2000, adamPRB02}. 
There is however a longstanding puzzle in the comparison between this theory
and experiment. 
The $g$-tensor distributions obtained from RMT 
are normalized to the average $(\langle g^2\rangle)^{1/2}$, which has to
be evaluated by independent arguments.
In the regime of
strong spin-orbit interaction, 
the two theoretical models\cite{matveev2000,adamPRB02}
predict $\langle g^2\rangle$
to have contributions from both spin and orbital magnetic moments
\begin{equation}
\langle g^2 \rangle = 
{3\over \pi\hbar}\tau_{\textrm{\small so}}\delta + \alpha {l\over R}\; ,
\label{av_g}
\end{equation}
where $l$ is the elastic mean free path, $R$ is the size of the particle
and $\alpha$ is a constant of order 1\footnote{In Ref.~\onlinecite{adamPRB02} 
Eq.~(\ref{av_g}) was derived
by including explicitly a phenomenological orbital term in the 
RMT Hamiltonian; whereas in Ref.~\onlinecite{matveev2000} this equation
was derived by expressing $\langle g^2\rangle$
and the energy absorption in a time-dependent magnetic field
in terms of the same phenomenological RMT parameter.}. 
In the limit of strong spin-orbit
scattering $\tau_{\textrm{\small so}}\delta \to 0$, 
the spin contribution vanishes and
only the orbital contribution 
$\alpha(l/R)\sim (\langle L_z \rangle)^2 $ survives.
The nanoparticles studied in Refs.~\onlinecite{petta2001, petta2002} 
are not disordered
and therefore $l\sim R$. Thus Eq.~(\ref{av_g}) 
predicts that $\langle g^2 \rangle$ should never be much less than 1.
In noble-metal nanoparticles, however, the measured values of 
$\langle g^2 \rangle$ are typically between 0.05 and 0.1.

RMT is a phenomenological approach which assumes that the statistical
properties of interest depend only on the symmetry of the Hamiltonian.
This assumption, although appealing and reasonable, is usually difficult
to justifies rigorously. This fact, together with the discrepancy between
theory\cite{matveev2000,adamPRB02} and experiment mentioned above, 
motivates the theoretical study of
$g$-factors in
metal nanoparticles presented here. Our investigation is
based on a semi-realistic microscopic tight-binding model
that we solve numerically.

From the comparison between RMT and our
microscopic theory we are able to shed some light on the discrepancy between
the experimental measurements
of $\langle g^2 \rangle$ and the value predicted 
in Refs.~\onlinecite{matveev2000,adamPRB02}.
The results of our model are in agreement with
most of the RMT conclusions
as to the functional form of the $g$-tensor distribution and its dependence
on the spin-orbit coupling strength. In particular
we find that $g$-tensors are strongly anisotropic when the nanoparticle
shape is not perfectly symmetric, and that this anisotropy originates from
mesoscopic quantum fluctuations of the nanoparticle wavefunctions.
However, we are able to go beyond RMT, in that our more detailed analysis
allows us to clarify the role played by the orbital motion and demonstrate
that its contribution to the average $g$-factor is very sensitive to the
character of the quasiparticle wavefunctions. Based on these results
we conclude that the small value of $\langle g^2 \rangle$ measured
experimentally can be understood if the orbital angular momentum 
of the tunneling states, in contrast to the RMT assumptions,
is still partially quenched.

The paper is organized as follows. In Sec.~\ref{model} we introduce
and discuss our model, with particular emphasis on the coupling
between orbital motion and external field. In Sec.~\ref{results}
we illustrate our numerical results, we compare them with RMT predictions,
and we discuss their implications
for the interpretation of the tunneling spectroscopy experiments.
A summary and concluding remarks are presented in Sec.~\ref{conclusions}.

\section{Model}
\label{model}

We model the nano-particle as a truncated fcc crystal lattice with
$N_a$ atoms. The shape of the system can be arbitrarily
varied to simulate the variability of realistic nanoparticles.
A $spd$ tight-binding-model is used with 18 orbitals at each 
atomic site, including the spin degrees of freedom.

The Hamiltonian, 

\begin{equation}
{\cal H}={\cal H}_{\rm band}+{\cal H}_{\textrm{\small so}}+{\cal H}_{\rm Zee},
\label{totham}
\end{equation}

has been introduced 
in a study of the quasiparticle properties
in ferromagnetic metal nanoparticles\cite{ac_cmc_ahm2002}. 
Here we give only a brief
description of the terms in Eq.(\ref{totham}).

The first term $H_{\rm band}$ is an orbital part, 

\begin{equation}
{\cal H}_{\rm band} = \sum_{i,j} \sum_{s} \sum_{\mu_1, \mu_2}
t^{i,j}_{\mu_1,\mu_2,s}
c^{\dagger}_{i,\mu_1,s}c^{\phantom{\dagger}}_{j,\mu_2,s}
\label{H1b}
\end{equation}

involving the Slater-Koster parameters~\cite{slater_koster}, 
$t^{i,j}_{\mu_1,\mu_2,s}$. Atomic sites are labeled by $i,j$, 
and $t^{i,j}$ couples up to  second nearest-neighbors.
The indices $\mu_1,\mu_2$ label the nine distinct atomic orbitals 
(one $4s$, three $4p$ and five $3d$). The spin degrees of
freedom, labeled by the index $s$, double the number of
orbitals at each site.

The second term describes a spin-orbit interaction, 
essentially atomic in character

\begin{equation}
{\cal H}_{\textrm{\small so}} = \xi\sum_i \sum_{\mu,\mu',s,s'}
\langle i,\mu,s| {\vec L}\cdot {\vec S}|i,\mu',s'\rangle
c^{\dagger}_{i,\mu,s}c^{\phantom{\dagger}}_{i,\mu',s'} ,
\end{equation}

reflecting the fact that relativistic effects are important
only when the electron is close to the nucleus. 
In spite of the local nature of these interaction,
the effect of ${\cal H}_{\textrm{\small so}}$ on the
spin-orbit relaxation time $\tau_{\textrm{\small so}}$,
compared to 
non-disordered infinite systems,
is strongly enhanced by the destruction of crystal symmetry 
due to the nanoparticle surface. This is, to a certain extent,
similar to the mechanism described by the Elliott relation.

A quantitative measure of the relative strength of the spin-orbit interaction 
is given by the dimensionless parameter $\lambda$\cite{brouwerPRL2000}

\begin{equation}
\lambda^2=\pi \frac{\hbar}{\tau_{\textrm{\small so}} \delta}\;.
\label{lambda2}
\end{equation}

The spin-orbit scattering is strong if $\lambda >> 1$ and weak 
if $\lambda << 1$.  
In the limit of weak spin-orbit interaction $\tau_{\textrm{\small so}}$
can be calculated perturbatively by Fermi golden rule. Here, however,
we use a more pragmatic approach: we {\it define} 
$\tau_{\textrm{\small so}}$ in terms of the average
spin-orbit quasiparticle energy shift\cite{ac_cmc_ahm2002}

\begin{equation}
\hbar \tau^{-1}_{\textrm{\small so}} 
\equiv \langle |\epsilon_n - \epsilon_n^0|\rangle,
\label{tauso}
\end{equation}

where $\epsilon_n$ and $\epsilon_n^0$ is the $n$-th 
eigenvalue with and without spin-orbit 
interaction respectively and 
the average $\langle\dots\rangle$ is performed over the spectrum
of the nanoparticle\footnote{In the limit of weak spin-orbit scattering,
this definition is equivalent to the Fermi golden rule result.}.

%
Since $\tau_{\textrm{\small so}}$ decreases weakly with particle size $L$, 
while $\delta$ varies as ${L}^{-3}$, the effective strength of the
spin-orbit interaction decreases with decreasing particle size.
In the experiments of Ref.~\onlinecite{petta2001, petta2002},  
containing up to several thousand atoms,
$\lambda$ can be as large as $10$ for Au nanoparticles.
In our theoretical studies we are able to deal numerically with nanoparticles
containing only up to a few hundred atoms, 
which would yield $\lambda << 1$ for a
realistic value of $\xi \approx 100$ meV.
In order to achieve larger
$\lambda$ values we therefore artificially increase the spin-orbit coupling 
strength $\xi$. For a disordered dot of a generic shape our numerical results
depend on the value of $\lambda$ but only weakly on the separate values
of $\xi$ and $\delta$.

The Zeeman part ${\cal H}_{ \rm Zee}$ in Eq.~(\ref{totham})
is conveniently divided into two 
terms,

\begin{equation}
{\cal H}_{\rm Zee} = -\mu_{\rm B}\sum_i \vec{B}\cdot 
\Big\{ \sum_{\mu,\mu',s,s'}\langle i,\mu,s|g_{s}\vec{S}+ \vec{L} 
|i,\mu',s'\rangle  c^{\dagger}_{i,\mu,s}c^{\phantom{\dagger}}_{i,\mu',s'} 
\Big\} + {\cal H}^{\rm NLOC}_{\rm Zee}
\label{zeeman}
\end{equation}

The first term of Eq.~(\ref{zeeman}) is the usual atomic 
contribution already considered in 
our earlier paper~\cite{ac_cmc_ahm2002}, arising from
the space dependence of the vector potential on an atomic length scale. 
In addition to this,
there is a second term,
a non-local orbital contribution arising from the magnetic flux encompassed
by closed loops describing the paths of an electron hopping from site to site. 
Only the latter term is accounted for in the RMT approach, whereas the former
term has a larger importance in many respects.
We discuss this contribution in detail in the next section.

\subsection{Non-Local Orbital Contribution (NLOC)}

The NLOC originates from delocalized electrons in an external magnetic 
field. The coupling to an external magnetic field in tight-binding 
theory is accomplished by the introduction of the Peierls 
phases~\cite{peierls_phases_1,peierls_phases_2,peierls_phases_3},

\begin{equation}
\exp(i\Theta_{i,j})=\exp \biggl [ \frac{ie}{ \hbar c} 
\int_{\vec{R}_i}^{\vec{R}_j} \vec{A}(x) \cdot d\vec{x}  \biggr]
		\approx \exp \biggl [ \frac{ie}{ \hbar c} 
		\frac{(\vec{R}_j-\vec{R}_i) \cdot 
		(\vec{A}(\vec{R}_j)+\vec{A}(\vec{R}_i))}{2}  \biggr] ,   
		\label{Peierls}
\end{equation}

modifying the Slater-Koster parameters so that,

\begin{eqnarray}
{\cal H}^{\rm NLOC}_{\rm Zee}= \sum_{i,j} \sum_{s} \sum_{\mu_1, \mu_2}
(\exp(i\Theta_{i,j})-1)t^{i,j}_{\mu_1,\mu_2,s} c^{\dagger}_{i,\mu_1,s}
c^{\phantom{\dagger}}_{j,\mu_2,s}.
\label{NLOC}
\end{eqnarray}

The non-trivial approximation of Eq.~(\ref{Peierls}), replacing the 
contour integral with a simple line integral has been justified 
by Ismail-Beigi {\it et al.}~\cite{lineint}.

Choosing the vector potential in the symmetric gauge,

\begin{equation}
\vec{A}=-\frac{1}{2} \vec{r} \times \vec{B},
\end{equation}

the Peierls phases can be conveniently rewritten in a form suitable
for perturbation theory purposes,

\begin{equation}
\Theta_{i,j}=-\frac{2 \pi}{4 \phi_0} \big [ (\vec{R}_j-\vec{R}_i)
\times(\vec{R}_i+\vec{R}_j) \big ] \cdot \vec{B}.
\end{equation}

The phase factor of Eq.~(\ref{Peierls}) is expanded  to first order 
in $\vec{B}$, 

\begin{equation}
\exp(i\Theta_{i,j})=1-i \mu_B \vec{L}^{NLOC} \cdot \vec{B} + O(B^2), 
\end{equation}

where $\vec{L}^{NLOC}$ is defined as

\begin{equation}
\langle i,\mu,s|\vec{L}^{NLOC}|j,\mu',s'\rangle=\frac{2 \pi}{4 \mu_B \phi_0} 
\big [ (\vec{R}_j-\vec{R}_i)\times(\vec{R}_i+\vec{R}_j) \big ]. 
\label{L}
\end{equation}

This expansion allows one to write Eq.~(\ref{zeeman}) 
to first order in (small) $\vec{B}$ in the familiar form

\begin{eqnarray}
{\cal H}_{\rm Zee} \approx -&\mu_B& \vec{\mu} \cdot \vec{B} \equiv  \\
-&\mu_{\rm B}&\sum_i \Big\{ \sum_{\mu,\mu',s,s'}
\langle i,\mu,s|g_{s}\vec{S}+ \vec{L}|i,\mu',s'\rangle 
c^{\dagger}_{i,\mu,s}c^{\phantom{\dagger}}_{i,\mu',s'} \Big\} 
\cdot \vec{B}\\
-&\mu_{\rm B}&\sum_{i,j} \Big\{ \sum_{\mu,\mu',s,s'}
\langle i,\mu,s|\vec{L}^{NLOC}|j,\mu',s'\rangle 
c^{\dagger}_{i,\mu,s}c^{\phantom{\dagger}}_{j,\mu',s'} \Big\} \cdot 
\vec{B}\; .
\label{Zpert}
\end{eqnarray}

We have explicitly checked that the Zeeman splitting due to the 
non-local interaction is gauge invariant in our numerical calculations,
although the definition presented above is obviously for a particular
gauge choice.

\subsection{Wannier states}

The basis set, $|j,\mu,s\rangle$, is to be interpreted as a set of
Wannier orbitals, 
extracted~\cite{slater_koster} from bulk properties. The Slater-Koster 
parameters, $t^{i,j}_{\mu_1,\mu_2,s}$, are Hamiltonian matrix elements 
of $|j,\mu,s\rangle$. Application of $t^{i,j}_{\mu_1,\mu_2,s}$ to a 
nano-particle  relies on their transferability, the assumption 
of local electronic environmental insensitivity to the boundaries of 
the system. 
The success of this procedure depends on the nature of 
the states involved . In noble metals the electronic density of the 
$sp$-bands is delocalized, and very different from a simple sum of 
atomic densities. The corresponding Wannier orbitals lose their 
atomic-like character in this limit, and the transferability is 
expected to be less robust, since the boundary conditions play an increasingly 
important role. The orbitals, $|j,\mu,s\rangle$, of $sp$-symmetry will 
thus have large hopping elements, a property expected to remain
the same in any transferable tight-binding treatment of these states. 
The $d$-orbitals originate from bands of a different nature, with rather 
localized electronic densities around the atomic sites. The 
corresponding Wannier orbitals remain atomic-like and 
hopping parameters are small. 

From these arguments it is immediately clear that the contribution from 
${\cal H}^{ \rm NLOC}_{\rm Zee}$ becomes increasingly important in the 
limit of extended Wannier states with large hopping parameters.

The eigenstates of this model are in general a mixture of  all basis 
orbitals, $|j,\mu,s\rangle$. An approximate separation of the eigenspectrum 
into bands of specific symmetries can still be made. 
For this purpose it is useful to define the 
projection operators, $P_s, P_p$ and $P_d$. 

\begin{eqnarray}
P_s &=& \sum_{i} \sum_{s} \sum_{\mu_s} |i,\mu_s,s \rangle  \langle i,\mu_s,s|\;, \\
P_p &=& \sum_{i} \sum_{s} \sum_{\mu_p} |i,\mu_p,s\rangle  \langle i,\mu_p,s|\;, \\
P_d &=& \sum_{i} \sum_{s} \sum_{\mu_d} |i,\mu_d,s\rangle  \langle i,\mu_d,s|\;, 
\label{proj}
\end{eqnarray}

with

\begin{equation}
P_s + P_p + P_d = 1\;.
\end{equation}
  
\begin{figure}
 \includegraphics{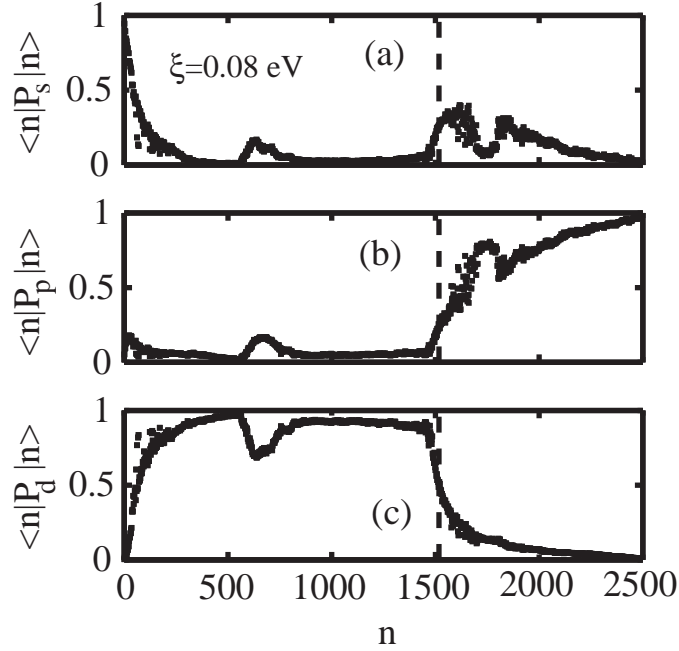}
 \caption{Matrix elements of $P_{\mu}$ according to Eq.~(\ref{proj}). 
  The eigenstate $|n>$ is for the Hamiltonian  of a 143-atom hemispherical 
  Au nanoparticle, with 5 atoms removed. The dashed vertical line marks the 
  Fermi level.}
 \label{norm}
 \end{figure}

These operators project onto the subspaces of $s$, $p$ or $d$-symmetry.
As an example, in Fig.~(\ref{norm}) 
we plot the expectation values of these projection operator
in the eigenstates $|n\rangle$ of the Hamiltonian of Eq.~(\ref{totham}). 
The calculations are done for
a 143 gold atom nanoparticle with hemispherical shape, from which 5 atoms
have been removed to break the rotational symmetry. 
It is seen that states in the middle of the occupied band have well-defined
$d$-character, with the exception of some states around $n=700$, which have
mixed $spd$-components and low-energy states, 
which have mainly $s$-character. High-energy states
far above the Fermi level have dominant $p$-character. States around the 
Fermi level are rather strongly mixed with the $d$-component decreasing
sharply with eigenvalue number.

%
\begin{table}

\begin{tabular}{||c|c|c|c|c|c||} \hline \hline

\emph{Data}           &      Material	 &     $<g_1>$      &     $<g_2>$    &    $<g_3>$     &      $\lambda$   \\ \hline

Exp.    	      &		Cu	 &  $0.9 \pm 0.3$   & $1.2 \pm 0.2$ & $1.6 \pm 0.3$  &	      $1.1 \pm 0.1$  \\ \hline

RMT	              & 	-	 & 	1.0	    &	   1.1      &	   1.6       &	      $1.1$  \\ \hline

Here	              &		Au	 &	1.1	    &	   1.3      &	   1.7       &	      $1.1$  \\ \hline

Exp.                  &         Cu       &  $0.4 \pm 0.2$   & $0.8 \pm 0.2$ & $1.3 \pm 0.3$   &	      $1.8 \pm 0.1$  \\ \hline

RMT	 	      &		-	 &	0.5	    &	   0.8      &	   1.3       &	      $1.8$  \\ \hline

Here	              &		Au	 &	0.7	    &	   0.9      &	   1.4       &	      $1.8$  \\ \hline \hline

\end{tabular}

\caption{Calculated mean values of $g_i$ compared to random matrix 
theory predictions and experiments. The $g$-factors of our model are 
based on d-states only. The experimental and random matrix theory 
data are from Petta and Ralph~\cite{petta2002}.}
\label{table}

\end{table}
%
%

\subsection{$g$-tensor}

In the absence of an external magnetic field, a nano-particle will have 
a doubly degenerate eigenspectrum, $\epsilon_{n}$, reflecting the 
formation of Kramer pairs in a time-reversal symmetric system. In the 
presence of spin-orbit interaction, the Zeeman splittings, 
$\epsilon_{n} \rightarrow \epsilon_{n} \pm \delta \epsilon_{n}$,
show anisotropic 
dependence on the direction of the applied field, $\vec{B}$. 
For small $\vec{B}$ 
this is conveniently described in terms of the $g$-tensor~\cite{slichter_PoMR},

\begin{eqnarray}
\delta \epsilon^{2}_{n}(\vec{B}) &=& \big ( \frac{\mu_{B}}{2} \big )^{2} 
     \vec{B} \cdot {\cal G}_{n} \cdot \vec{B} \\
	   {\cal G}_{n} &=& G^{T}_{n}G_{n}.
\end{eqnarray}

By using standard degenerate perturbation theory in ${\cal H}_{\rm Zee}$,
the $G_{n}$ matrices can be related to the matrix elements of the 
(dimensionless) magnetic moment operator $\vec \mu$, 

 \begin{eqnarray}
(G_n)_{1,j} + i (G_n)_{2,j} &=& -2 \langle T \psi_n | \mu_j | \psi_n \rangle \\
		(G_n)_{3,j} &=& 2 \langle \psi_n | \mu_j | \psi_n \rangle\; .
\end{eqnarray}

Here $|\psi_n\rangle$ and $|T \psi_n\rangle$ are the
time-reversed pair of eigenstates of ${\cal H}$ in the absence 
of $\vec{B}$, corresponding
to the eigenvalue $\epsilon^{2}_{n}$.
The diagonalization of ${\cal G}_{n}$, or in other words a suitable choice of 
coordinate system for the external field $\vec{B}$,  allows one to write
$\delta \epsilon_{n}(\vec{B})$ in terms of the 3 eigenvalues 
$g_j^2$, $j=1,2,3$, of ${\cal G}$

\begin{eqnarray}
\delta \epsilon_{n}(\vec{B}) =  \frac{\mu_{B}}{2}
	\sqrt{   g_1^2 B_1^2 +  g_2^2 B_2^2  +   g_3^2 B_3^2    },
\end{eqnarray}

where $B_j$, $j=1,2,3$, are the components of the magnetic field along
the ``principal axis'',
$\hat{e}_1, \hat{e}_2, \hat{e}_3$, 
defined by the three normalized eigenstates of ${\cal G}$. 
We refer to $g_j, j=1,2,3$ as the principal  $g$-factors\cite{brouwerPRL2000}.
The Zeeman splittings, 
$\delta \epsilon_{n}(\vec{B})$, can also be obtained from direct
numerical diagonalization of the Hamiltonian for a given direction of 
$\vec{B}$. This approach is of course time demanding, but served 
as a check of the correctness of the perturbative approach, 
since the two methods should give identical results for small $\vec{B}$. 

In the limit of zero spin-orbit interactions the three principal $g$-factors
are identically equal to 2.
In this case the Kramer pairs are pure spin states 
and the angular momentum is 
quenched. With the introduction of spin-orbit interaction spin characters 
mix, and non-zero orbital contributions~\cite{orbital-contr} to $g_i$ are 
expected. At weak spin orbit strength the spin character mixing is weak 
and the orbital contribution to $g_i$ is small. For large $\lambda$
the spin character mixing is strong, while orbital contributions are 
expected to increase.

\section{Results and Discussion} 
\label{results}
\begin{figure}
 \includegraphics{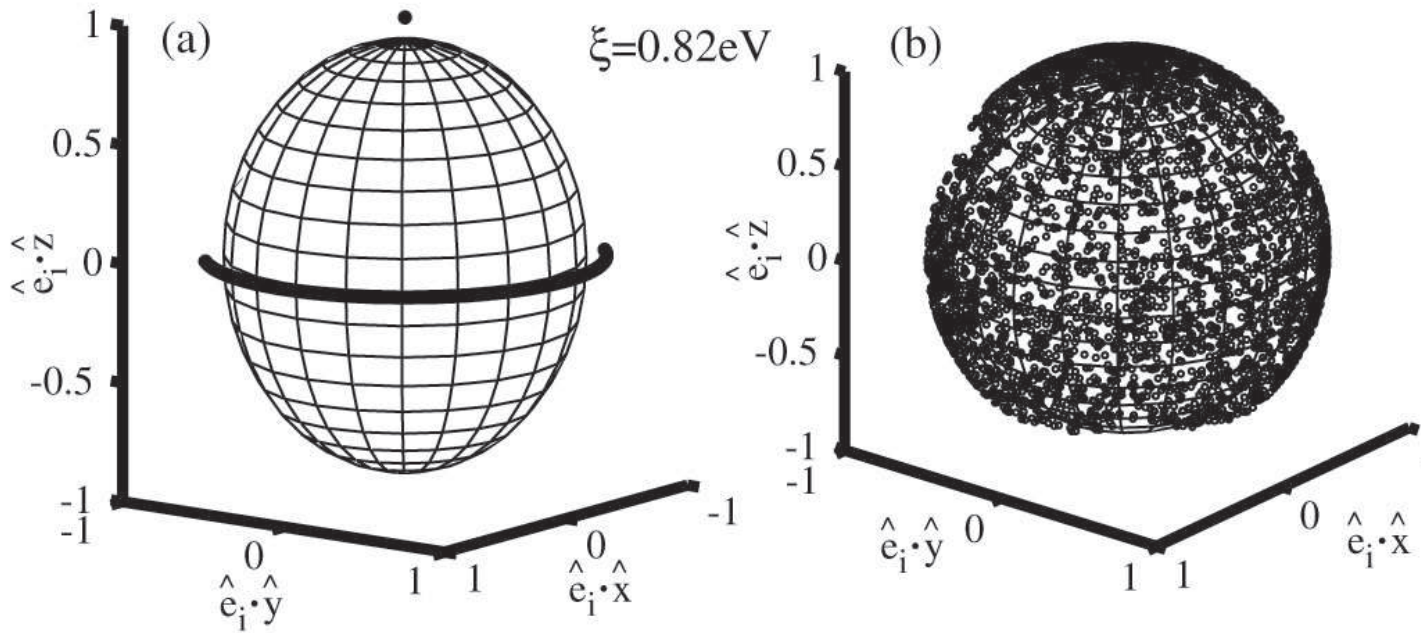}
 \caption{Principal axis directions on the unit sphere for a 
 	143 atom Au hemisphere in (a). In (b) 
 	5 atoms have been removed from the hemisphere. Each point represents
	one principal axis direction.}
 \label{prin}
 \end{figure}

In this section we present results of numerical calculations of $g$-tensors
for $Au$-nanoparticles containing 143 atoms. Similar results, not shown here,
were obtained for larger and smaller sizes.
We first consider the distribution of principal-axis directions.
Fig.(\ref{prin}a) shows calculated principal axis directions
for a perfect hemisphere. The direction of
$\hat{e}_i$ is strongly dependent on the symmetry of the nanoparticle,
with one axis always along the normal of the base 
plane. Due to the crystal fcc symmetry and the hemispherical truncation, 
the two remaining $g$-factors will be degenerate and two principal axes
can be arbitrarily chosen within the base plane. This is shown in 
Fig.(\ref{prin}a), where we plot the direction of the principal axis directions
on the unit sphere for all the quasiparticle orbitals of the nanoparticle.
The situation changes dramatically with a small distortion of the 
shape obtained by removing five atoms 
from the hemispherical truncation. This slight breaking of the spherical
symmetry produces a completely random distribution of principal axis,
no longer preferentially aligned along crystalline or shape directions,
as clearly seen in Fig.~(\ref{prin}b).
A random distribution of the principal axis directions was originally
predicted by RMT\cite{brouwerPRL2000}. 
Recent experiments~\cite{petta2002} measuring principal $g$-factors  
find random spatial orientation of the corresponding axis directions, 
$\hat{e}_i$. The measurements were performed on
Cu nanoparticles with approximately hemispherical shape.
The results demonstrates that small irregularities in the 
boundaries in an otherwise ordered sample are enough to produce
completely randomized directions, exactly as happens in our model.

 \begin{figure}
 \includegraphics{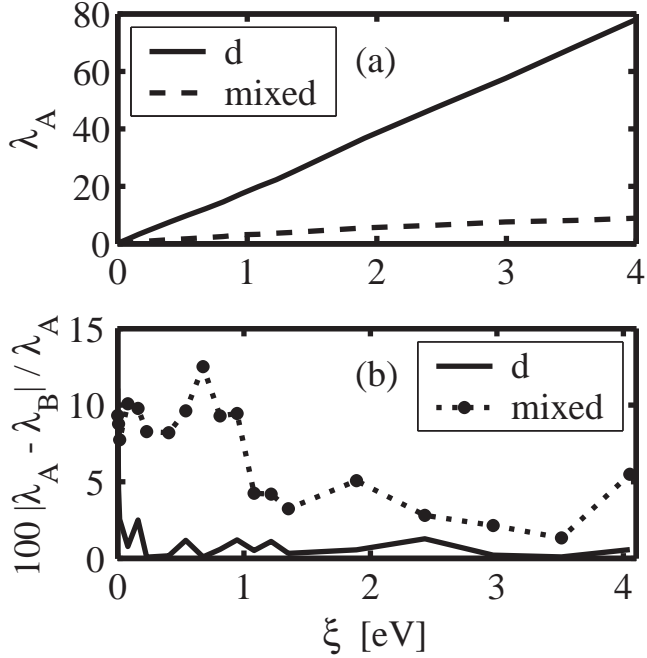}
 \caption{ $\lambda_{A}(\xi)$ in (a) shows the spin-orbit strength for a 
 143-atom Au nanoparticle with 5 atoms removed. $\lambda_{B}(\xi)$ originates 
 from a nanoparticle with a strongly disordered shape, but the same $N_a$
 as A. The states are divided into $d$-states, $<n|P_d|n> > 0.7$, 
 and mixed states, $<n|P_d|n> < 0.7$ . (b) shows the differences 
 in $\lambda$ between the two systems in percent. }
 \label{shape}
 \end{figure}

Once the perfect axial symmetry of the  hemispherical nanoparticle 
is broken, for example by removing a few atoms, we find that 
the numerical results depend
weakly on the particle shape: nanoparticles with shapes that are even less
symmetric display qualitatively similar distributions of 
principal $g$-factors.
It turns out that the dimensionless spin-orbit strength
$\lambda$, defined in Eqs.~(\ref{lambda2})-(\ref{tauso}),
is the crucial parameter that controls the distributions.
A  meaningful evaluation of $\lambda(\xi)$ for a given value of $\xi$ is
however made complicated by the fact that states of different orbital character,
as determined by the projection operators discussed above, yield 
very different values of $\lambda$ for the same $\xi$.
Therefore we divide the eigenstates into two groups of states,
which we refer to $d$- and {\it mixed} states. An eigenstate is
operationally considered a $d$-state if $\langle n|P_d|n\rangle >0.7$;
it is a  mixed state otherwise. 
We will use this distinction extensively below.
For a given value of $\xi$, the parameter $\lambda$ is then calculated
separately for these two group of states. This procedure turns out to be
very useful in the interpretation of our numerical results. A first
example of this is shown in Fig.~(\ref{shape}). In Fig.~(\ref{shape}a) 
we plot $\lambda(\xi)$ as a function of $\xi$ for these two groups of states
for a given nanoparticle. In both cases $\lambda$ increases 
approximately linearly with $\xi$, but the slope is much larger for the
$d$-states. In Fig.~(\ref{shape}b) we plot relative differences of
the $\lambda$ for two nanoparticles with the same number of atoms
($N_a =143$) but very different shapes, again separating $d$- and mixed
states. Here $\lambda_A$ refers to a nanoparticle with 
hemispherical shape with 5 atoms removed;  $\lambda_B$ refers to a
nanoparticle with a strongly disordered shape. It is seen that 
$\lambda$ for the $d$ states is essentially the same for both nanoparticles
for all values of $\xi$. On the other hand mixed states give a value
of $\lambda$ that is more sensitive on the shape of the nanoparticle,
although the relative difference is always less than $10\%$.
It follows that the two nanoparticles will have very similar $g$-factor
distributions, with a stronger shape dependence of the $g$-factor of
mixed character states.

We now come to the discussion of the main results of our paper and to
the comparison with RMT predictions. A very useful quantity for 
this comparison is 
the average of the sum of the squares of the principal $g$-factors
\begin{equation}
\langle g^2 \rangle = \frac{1}{3}\langle g_1^2 + g_2^2+ g_3^2\rangle\;.
\end{equation}

 \begin{figure}
 \includegraphics{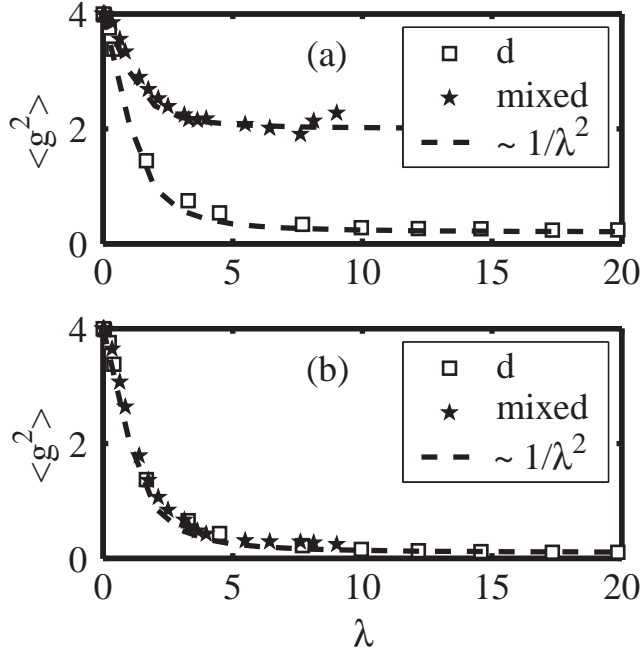}
 \caption{Mean square principal $g$-factors of a 143 atom Au hemisphere 
  with 5 atoms removed. The states are divided into $d$-states, 
$<n|P_d|n> > 0.7$ ,
  and mixed states, $<n|P_d|n> < 0.7$ . 
The full Hamiltonian is used in (a) while
  the NLOC to the Zeeman term is lacking in (b). The fitting function 
  is given in Eq.~(\ref{fitting})	}
 \label{gsquare}
 \end{figure}

In Fig.~\ref{gsquare} we plot $ \langle g^2 \rangle$ as a function of
$\lambda$ for a Au hemispherical nanoparticle with 5 atoms removed,
separating again the contribution from $d$- and mixed states.
The results in Fig.~\ref{gsquare}(a) are for the case in which
the spin term and both orbital terms are included in 
the magnetic moment of Eq.~(\ref{mmoment}).
As expected, $ \langle g^2 \rangle \to 4$ in the limit $\lambda \to 0$
of zero spin-orbit
coupling and decreases monotonically with increasing $\lambda$.
In the limit of strong spin-orbit coupling, $\lambda>>1$,  
$\langle g^2 \rangle$ tends
to a constant value, which is relatively small ($\approx 0.3$)
for $d$-states but is of order 2 for mixed states. This saturation value
in the strong spin-orbit scattering limit comes almost entirely from the 
non-local orbital contribution (NLOC) to the magnetic moment.
Indeed Fig.~\ref{gsquare}(b)
shows that that the saturation value is negligible when the NLOC 
is not included; in this case the $\langle g^2 \rangle$ vs. $\lambda$ for
$d$- and mixed states fall on the same curve
which goes to zero like $1/\lambda^2$ for large
$\lambda$. Note that the large saturation value for mixed states,
when the NLOC is present, is consistent with the fact that $d$-states
are linear combination of atomic orbitals with small hopping  parameters,
whereas mixed states tend to be more free-electron like.

We found that the calculated 
$\langle g^2 \rangle ( \lambda )$ could always be fit to

\begin{equation}
\langle g^2 \rangle(\lambda)=\frac{4-\nu(\lambda)}{1+\lambda^2} + 6\eta^2
\label{fitting}
\end{equation}
in the entire range of $\lambda$. Here $\eta$ is a constant independent
of $\lambda$, whereas $\nu(\lambda)$ is a function weakly dependent on
$\lambda$ such that $\nu(\lambda=0) = 6\eta^2$ and $\nu(\lambda)\to 1$ for
$\lambda >>1$. In practice a reasonable fit (shown in the figure)
is obtained by fixing $\nu= 6\eta^2$. Remembering the relationship 
between $\lambda$ and $\tau_{\textrm{\small so}}$ given in Eq.~(\ref{lambda2}),
we can see that Eq.~(\ref{fitting}) is 
perfectly consistent with the RMT prediction\cite{matveev2000, adamPRB02} 
in the strong spin-orbit
scattering limit given in Eq.~(\ref{av_g}). The first term in Eq.~(\ref{fitting})
comes essentially from the spin contribution to the magnetic moment. The
second term originates from orbital contributions and it has been written
in this specific way to make contact with the the RMT phenomenological
parameter $\eta$ that describes its coupling of the 
magnetic field to the orbital
angular momentum\cite{adamPRB02}. The RMT parameter $\eta^2$ can be estimated
for specific physical systems
by computing $\langle n| L_z^2|n\rangle$ in the absence of spin-orbit
coupling\cite{matveev2000,adamPRB02}. For a ballistic sphere with
diffusive boundary conditions one gets 
$\eta^2 =1/8$\cite{matveev2000,adamPRB02}. Thus for no disorder RMT predicts
for $\langle g^2 \rangle$ a saturation value $\approx 1$,  which according to
our model would correspond to the case of mixed states.

\begin{figure}
 \includegraphics{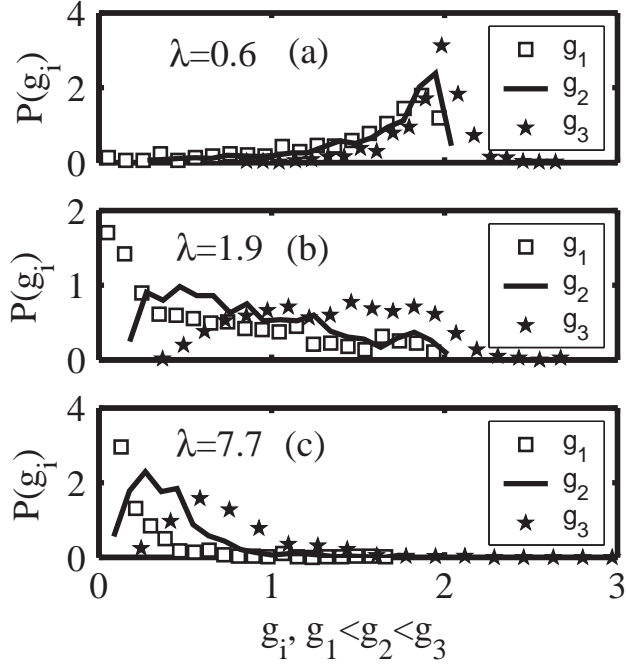}
 \caption{$g$-factor distributions of $d$-states for three different values of 
  $\lambda$. The system is a 143 atom
 Au hemisphere with 5 atoms removed.}
 \label{RMTdist}
 \end{figure}

\begin{figure}
 \includegraphics{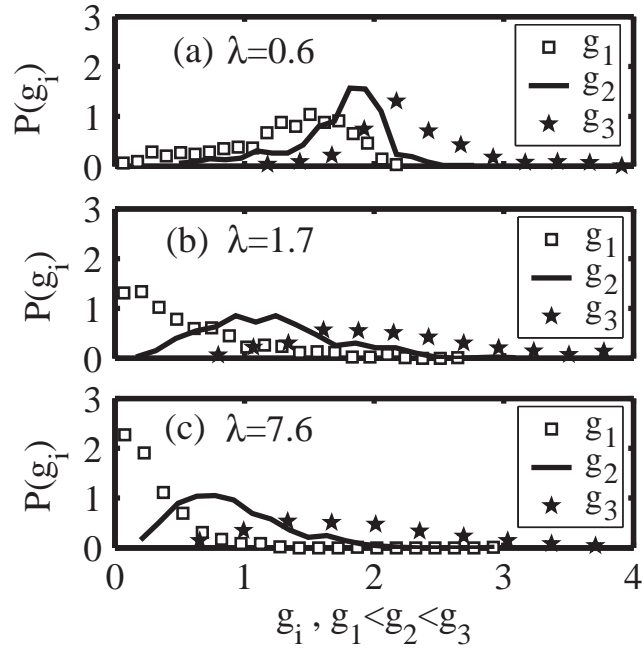}
 \caption{$g$-factor distributions of mixed states for three different 
 values of $\lambda$. The system is a 143 atom
 Au hemisphere with 5 atoms removed.}
 \label{RMTdistMIX}
 \end{figure}

We can further analyze the relationship between RMT and our microscopic
model by comparing the distribution of the three individual
principal $g$-factors
in different regimes of the spin-orbit coupling strength.
In Fig.~\ref{RMTdist} and Fig.~\ref{RMTdistMIX} we plot 
the distribution of $g_1$, 
$g_2$, and $g_3$ for $d$ and mixed states respectively. Translating
into RMT language,  Fig.~\ref{RMTdist} and Fig.~\ref{RMTdistMIX} 
correspond to small and
large orbital contribution respectively. The sub-cases 
(a), (b), (c) in each figure correspond to the regimes of weak, 
intermediate and strong spin-orbit
scattering respectively. The values of $\lambda$ are chosen to compare with
the cases considered in Refs.~\onlinecite{brouwerPRL2000, adamPRB02}. 
For weak spin-orbit interaction,
we find
that typically $g_1 \approx g_2 < 2$ and $g_3 \approx2 $ when
the spin contribution to the $g$ tensor dominates (i.e. $\eta << \lambda$
as in Fig.\ref{RMTdist}(a) ).
On the other hand, when the orbital
contribution dominates the $g$-tensor ($\eta >> \lambda$
as in Fig.\ref{RMTdistMIX}(a) ), $g_1 < 1$, $g_2 \approx 2$, $g_3 >2$.
Both cases are in remarkable agreement with 
RMT predictions\cite{brouwerPRL2000, adamPRB02}

The trends of our numerical distributions for the cases of intermediate 
and strong spin-orbit scattering also agree with the
RMT scenario\cite{brouwerPRL2000}. 
In particular in the strong spin-orbit scattering regime and for weak
orbital contribution (see Fig.~\ref{RMTdist}(c) ), 
all three principal $g$-factors are peaked at values
smaller than 0.5. In this case, typical values of $g$ are always much
smaller than the free electron value $g_0= 2$.

We conclude this section by attempting a comparison between the results of
our model and the experimental
results of Ref.~\onlinecite{petta2002}.
In Table~\ref{table} we have summarized: (i) the experimental measurements of
the average values of $g_i, i =1,2,3$,  for two
Cu nanoparticles--raws labelled by ``Exp''; 
(ii) the RMT results, including spin contribution only--raws 
labelled by ``RMT'';
and (iii) the results of our calculations for $d$-states,
including both spin and orbital contributions
--raws labelled by ``Here''\footnote{The calculations
presented here are done on gold nanoparticles. 
Gold was one of the materials used in 
earlier $g$-factor experiments\cite{petta2001}, 
which showed the same qualitative behavior for all noble-metal nanoparticles.
Since all noble metals have similar electronic
structures at the Fermi level, we expect our model to produce similar results 
for all them (Au, Ag and Cu), provided that we compare noparticles with the same
relative spin-orbit interaction strength. As we explained in Sec.\ref{model}, 
this strength is controlled by the parameter $\lambda$, which depends on the
intrinsic spin-orbit coupling strength $\xi$ and the volume-dependent 
single-particle mean-level
spacing. Here we use $\xi$ as a free parameter to generate different values
of $\lambda$ and 
mimick particles of different sizes.}.
The procedure
to extract these numbers is the following.
First the value of $\lambda$ corresponding to a particular nanoparticle is
obtained\cite{petta2002} by matching the RMT value of $\langle g^2\rangle$
with the same quantity measured experimentally. Given this $\lambda$,
the average values of the principal $g$-factors $\langle g_1 \rangle$,
$\langle g_2 \rangle$, and $\langle g_3 \rangle$ can then be calculated
theoretically by RMT. This was done in Ref.~\onlinecite{petta2002}
and the results are reported in Table I.
We can also compute the averages of $g_i$ within our microscopic model
by choosing the parameter $\xi$ for a given nanoparticle
so that the value of $\lambda$ obtained from Eqs.~\ref{lambda2}-\ref{tauso}
is equal to the value extracted from the experiment using RMT.
The averages predicted by the two theories (RMT and our macroscopic
model)
can then be compared with each other and with 
the values measured experimentally.

From the table one can see that RMT -- 
{\it with orbital contributions neglected} -- predicts
average values of $g_i$ in good agreement with the experiments for two
different $\lambda$'s  corresponding to two different nanoparticles. 
The message that we want
to convey here is that the average $g$ values of $d$-states
predicted by our
microscopic model, at the same nominal values of $\lambda$ defined 
through Eqs.~\ref{lambda2}-\ref{tauso}, 
are also in good agreement with RMT and the experiments, 
{\it even when the orbital contributions are included}\footnote{In fact,
RMT seems to agree even better with experiment than our theory.
One possible reason is that the definition of $\lambda$ used in our
model -- see Eqs.~\ref{lambda2}-\ref{tauso}-- 
might differ by a constant factor 
from the RMR $\lambda$. Specifically
in the two cases considered in Table I, our model would give a better
agreement with experiment if we used a slightly smaller $\lambda$.}.
Thus our resuls suggests a scenario, namely $g$-factor distributions
of states of mainly $d$ character,
which, if realized, would solve the puzzle posed by the comparison
between RMT and experiment. Note that RMT is based on a
picture of single-particle wavefunctions essentially ergodic in space
and it is therefore unable to capture the nature of wavefunctions more localized
around atomic cores or defects, which instead are naturally described
within our tight binding model.

The problem with the scenario proposed here 
is that the Fermi level for noble-metal
nanoparticles lies near states of both $d$-like and mixed-like character,
(see Fig.~\ref{norm} )
and therefore the latter ones most likely play a very 
relevant role in the tunneling
experiments.
However if we use the procedure described above and compute
$g$-factor averages including mixed states,
we find values that are too large in comparison with experiment. 

As possible solutions, 
we consider the following: 
(i) electrons are, in fact, tunneling
into states of pure $d$ character. (ii) Electrons are tunneling into states
of mixed characters, but the nature of these delocalized states is 
strongly affected by the surface and therefore
their characteristics, including their orbital angular momentum,
are profoundly modified. This is an effect which is clearly not included
in our model, and most likely it is more important for mixed states
than for $d$-states. (iii) electron-electron interaction beyond the
simple mean-field approximation incorporated in the SK parameters could also
modify the electronic states and partially quench 
their orbital angular momentum.
This effect is also not included in our model.
Note that recent studies\cite{ee_int_size} 
have demonstrated a sharp increase
of electron-electron interaction due to surface induced reduction of
screening. 
At the moment we find the occurrence of
explanation (i) problematic: apart from the fact that in our calculations
all states of pure $d$-character are below the Fermi level and therefore
not available for tunneling, states of $s$ character have much
larger tunneling probabilities than $d$-states,
when (as in the present case)
the barriers separating the grain from the electrodes are made
of aluminum oxide. 
Explanations (ii) or (iii) or a combination of the two seem more compelling
and alluring.
However pursuing these lines requires theoretical 
modeling beyond the scopes of the
present work.

\section{Conclusions} 
\label{conclusions}

In summary, we have presented a theoretical study of the statistical properties
of $g$-tensors for individual quasiparticle energy levels of 
metal nanoparticles, based on numerical calculations for a 
simplified but realistic model
that treats spin-orbit interactions microscopically.
Our theory of the $g$-tensors includes
both spin and orbital contributions to 
the magnetic moment. We have shown that even small deviations from a perfectly
symmetric nanoparticle shape cause random fluctuations in the quasiparticle
wavefunctions, which,  in turn, are the source 
of strong anisotropies of the $g$-tensors 
and strong level-to-level fluctuations both in the principal $g$-factor
values and in the directions of their principal axis. A dimensionless parameter
measuring the strength of the spin-coupling controls the $g$-tensor
distributions for nanoparticles of generic shape, in excellent agreement with
the prediction of random matrix theory.
Our work sheds light on the relative importance of the spin vs
orbital contributions to the $g$-factors and the strong dependence of the
latter on the orbital character of the wave-functions.
The presence near the Fermi energy of states of both $d$-like 
and $sp$-like character is responsible for aspects of the $g$-factor physics
that are not captured by Random Matrix Theory.
The small values of $g$-factors measured experimentally
suggest that the orbital angular momentum of the tunneling states near the Fermi energy is most
likely still partially quenched, even in the presence of 
strong spin-orbit scattering.
Our calculations demonstrate that angular momentum quenching cannot be due simply to 
the strong $d$-orbital hybridization of states near the Fermi energy.   
Non-trivial changes in electronic structure might be responsible, perhaps due 
to irregularly shaped nanoparticle boundaries that produce orbitals more localized 
at the surface than those in the model that we have studied.  
Enhanced correlation effects near the surface, could also play a role.  
If surface imperfections are the main source of orbital localization,
large $g$-factors should be observable in nanoparticles with very 
regularly shaped boundaries in which all orbitals are extended across the entire particle.

\section{Acknowledgments}
We would like to thank Dan Ralph for several
explanations of his experimental results,
and Piet Brouwer, Mandar Deshmukh
and Leonid Glazman for stimulating discussions.
This work was supported in part by the Swedish Research Council
under Grant No:621-2001-2357, by the faculty of natural science
of Kalmar University,
and in part by the National Science Foundation
under Grants DMR 0115947 and DMR 0210383.
Support from the Office of Naval Research under Grant N00014-02-1-0813
is also gratefully acknowledged.


\end{document}